\documentclass[journal,11pt]{IEEEtran}
\usepackage{amsmath}
\usepackage{booktabs}
\usepackage{graphicx}
\usepackage{cite}
\usepackage[hyphens]{url}
\usepackage{array}
\usepackage{tikz}
\usetikzlibrary{positioning,arrows.meta}
\graphicspath{{../figures/}}
\begin{document}
\title{Idempotency Is Not Recovery: A Fault-Injection Study of Language-Model Agent Tool Actions}
\author{Song~Xu\thanks{Song Xu is with the University of Electronic Science and Technology of China, Chengdu, China.} \thanks{Preprint; not peer reviewed. Licensed under the Creative Commons Attribution-NonCommercial 4.0 International license.} \thanks{This is an earlier single-author preprint. A later revision with additional authors has been submitted to IEEE Transactions on Reliability for consideration. This work has been submitted to the IEEE for possible publication. Copyright may be transferred without notice, after which this version may no longer be accessible.}}
\maketitle

\begin{abstract}
When a language-model agent's tool call is interrupted, the action may already have changed an external system without confirmation. Retrying risks duplication; assuming success may conceal an action that never occurred. We study these failure modes in a controlled agent runtime combining authorization, an action ledger, target-side idempotency, reconciliation, and evidence sealing. We inject ten single-host fault classes and compare four configurations and three ablations using a target-side oracle separate from the runtime ledger. On each of Windows and Linux, 7,513 trials used 96 captured model responses from 24 constructed cases. In 200 known original crash and replay pairs, the full configuration recorded 0/200 repeated effects versus 117/200 for each simpler baseline. A ledger-related ablation had 170/170 orphaned effects versus 0/170 and 200/300 recovery successes versus 300/300 for the full configuration. Disabling reconciliation left 100/100 known post-effect crash outcomes ambiguous despite no observed duplicates. A separate supplement designed after the initial results and reusing the same inputs found a similar pattern under response loss or delay. The full runtime had about one-seventh of fail-open throughput, while injected evidence-write faults left 75/100 trials per platform unresolved. These findings distinguish duplicate suppression from recovery and identify their measured reliability and performance tradeoffs within the tested benchmark.
\end{abstract}


\section*{Abbreviations and Nomenclature}
\noindent\textbf{B0--B3:} fail-open, fail-closed, durable-authorization, and full runtime configurations. \textbf{A1--A3:} ablations of the ledger-related bundle, reconciliation, and evidence sealing. \textbf{OP1--OP3:} alert append, ticket upsert, and conditionally guarded case transition.

\noindent\textbf{F1--F5:} invalid authorization; crash after reservation; crash before effect; response loss or delay after effect; and crash after effect before completion. \textbf{F6--F10:} capture I/O failure; completion failure; pause/resume or stop; replay; and evidence tampering. \textbf{RQ1--RQ5:} authorization; effect integrity; tamper detection; recovery and overhead; and cross-platform reproducibility.

\section{Introduction}
An agent can change an external system and never learn that it did. Suppose a language-model agent updates a support ticket and the response is lost. Retrying may apply the update twice; treating the silent call as successful may hide an update that never happened; stopping leaves no one certain which is true. As language-model agents move from generating answers to invoking tools that change records, tickets, and application state, this gap between what happened at the target and what the agent knows becomes a reliability problem in its own right.

Agent benchmarks and systems research address complementary parts of this problem. Agent benchmarks evaluate whether a model chooses and invokes the right tool, follows policy, or resists misuse \cite{liu2024agentbench,yao2025taubench,ruan2024toolemu,debenedetti2024agentdojo}; a correct proposal, however, does not determine what the target records after a crash or a lost response. Systems research supplies the remedies---write-ahead logging, stable request identities, idempotency, and recovery \cite{mohan1992aries,garciamolina1987sagas,lee2015rifl}---and the end-to-end argument places the final check at the endpoints \cite{saltzer1984endtoend}. Recent agent-reliability studies monitor worker liveness or inject faults into agent--model interactions \cite{chen2026reliharness,bateman2026aes}. A complementary question is how tested safeguards and ablation bundles relate to target-side outcomes and measured overhead in a concrete agent runtime.

We provide that measurement for one implemented runtime. Its full configuration combines fail-closed authorization, a durable action ledger, stable action identities with target-side idempotency, post-crash reconciliation, and sealed evidence. We inject ten classes of single-host faults, including process terminations and I/O exceptions on real code paths, and compare four execution configurations and three ablations against a target-side oracle recorded separately from the runtime ledger. The inputs are 96 previously captured model responses from 24 constructed cases (42 distinct normalized proposals), replayed identically across configurations; on each of Windows and Linux, 7,513 trials were recorded and recomputed.

Our main findings are as follows.
\begin{itemize}
\item \emph{Fail-closed authorization blocks the tested unauthorized effects.} Rejecting invalid authority before dispatch blocked every injected unauthorized action (0/100 versus 100/100 for fail-open). The simplest fail-closed configuration matched the full runtime on this outcome; its measured operation-window throughput was not lower than the fail-open baseline.
\item \emph{Duplicate suppression and recovery are distinct.} In the 200 known crash and replay pairs, the full configuration recorded 0/200 repeated effects versus 117/200 for each simpler baseline; the ledger-bundle ablation recorded 0/300.
\item \emph{Idempotency is not recovery.} Without the ledger bundle, 170/170 effects were orphaned and recovery fell from 300/300 to 200/300. With the ledger but without reconciliation, every known crash-after-effect pair remained ambiguous (100/100), although no duplicates occurred.
\item \emph{Durable configurations incurred measured overhead.} The durable-authorization and logging configuration had roughly half the measured throughput of the volatile fail-closed configuration; their pair-count fields matched in nine comparisons against the full configuration. The full runtime ran at about one-seventh of fail-open throughput, and because it would not finalize without evidence, injected evidence-write faults left 75/100 trials unresolved.
\end{itemize}
A supplement designed after the original outcomes and reusing the same inputs showed the same duplicate and reconciliation patterns in observed trials under verified response loss or delay: repeated effects in 0/100 versus 66/100 trials, and every observed reconciliation-disabled trial ambiguous.

This paper makes three contributions:
\begin{enumerate}
\item A fault model and effect oracle for an agent's tool-action boundary: ten fault classes at declared cutpoints, and a target-side oracle that separates what happened from what the runtime recorded.
\item A paired, cross-platform ablation study comparing failure outcomes across safeguard configurations and bundles, reporting 44 pre-specified comparisons per platform, including incomplete ones.
\item A quantified account of what these safeguards cost in throughput, latency, storage, and availability under evidence faults, translated into design implications for agent runtimes.
\end{enumerate}
Five pre-specified research questions (RQ1--RQ5) organize the evaluation, and Table~\ref{tab:rq} answers them. Section~\ref{sec:results} separates three evidence layers: the original pre-specified analysis, the later response-loss supplement, and post-hoc sensitivity summaries. All inputs are reused across configurations and platforms, so the two platforms show consistent behavior of one implementation rather than independent replication; Section~\ref{sec:threats} discusses this and other limits.

\section{Background and Related Work}
\subsection{Recovery, idempotency, and authenticated records}
Write-ahead logging and recovery protocols establish how durable records can support database recovery \cite{mohan1992aries}; sagas address long-running work through decomposed actions and compensations \cite{garciamolina1987sagas}. RIFL records request identities and results to provide exactly-once semantics for its storage setting \cite{lee2015rifl}; Helland discusses the role of stable identities in retry-safe operations \cite{helland2012idempotence}. Beldi, AFT, Durable Functions, and Boki study persistent execution or transactional semantics in serverless settings \cite{zhang2020beldi,sreekanti2020aft,burckhardt2021durable,jia2021boki}. Our single-host study does not inherit their distributed guarantees.

Authenticated audit logging and tamper-evident data structures have also been extensively studied \cite{schneier1999audit,crosby2009tamper}. Our capture history uses a symmetric message authentication code and a hash chain with a controlled test key; it is not a proof against an attacker who controls that key. The contribution is an empirical comparison of a particular combination of existing mechanisms at a tool-action boundary, not a new cryptographic primitive or general exactly-once theorem.

\subsection{Failure testing and agent evaluation}
Fault-injection methodology and surveys describe how injected faults, activation, and observations support dependability assessment \cite{arlat1990fault,hsueh1997fault,natella2016survey}. Chaos engineering applies controlled disruptions to services \cite{basiri2016chaos}, and studies of production failures motivate checking error-handling paths \cite{yuan2014testing}. FATE and DESTINI exercise cloud recovery \cite{gunawi2011fate}, lineage-driven injection selects faults from data dependencies \cite{alvaro2015lineage}, CHESS explores concurrent schedules \cite{musuvathi2008chess}, and bounded crash testing examines file-system consistency \cite{mohan2018crash}. Our harness is narrower: it targets declared points in a single-host agent runtime and a controlled SQLite target. Process terminations and injected I/O exceptions exercise real code paths, but are not equivalent to arbitrary power-loss or hardware-failure experiments.

AgentBench evaluates agent behavior across interactive environments \cite{liu2024agentbench}. $\tau$-bench measures tool-agent-user interaction and policy-following reliability \cite{yao2025taubench}; ToolSandbox adds stateful tool execution \cite{lu2025toolsandbox}; ToolEmu uses an emulated tool environment for risk assessment \cite{ruan2024toolemu}; and AgentDojo and AgentHarm examine agent misuse and attacks \cite{debenedetti2024agentdojo,andriushchenko2025agentharm}. Our work freezes model proposals before the fault trials and measures target effects and runtime state, not model reasoning quality, prompt-injection robustness, or live provider availability.

AgentSpec enforces runtime rules \cite{wang2026agentspec}, and ToolMaze evaluates replanning under tool perturbations \cite{zhu2026toolmaze}. ReliHarness monitors worker liveness and records execution evidence \cite{chen2026reliharness}; we instead assess target effects and post-fault reconciliation. An effect-history model classifies tool-boundary anomalies \cite{trofimov2026toolboundary}; we empirically examine one controlled boundary rather than prove a general model. Agent Error Simulator injects faults into agent--model interactions \cite{bateman2026aes}; we hold proposals fixed and examine target effects and recovery. ReliabilityBench also studies controlled agent failures \cite{gupta2026reliabilitybench}. We do not compare against these systems directly. Our contribution is narrower and complementary: for one runtime, an evaluation of how tested safeguard configurations and bundles relate to target-side failures and measured overhead.

The research repository descends from Nexa, whose broader hierarchical-context and bridge-scheduling architecture is described in a related patent \cite{xu2026nexa}. The experiments evaluate a specialized single-host tool-action runtime, not the complete scheduler, context hierarchy, or learning loop.

\section{Runtime and Mechanisms}
\subsection{Execution boundary}
Each captured proposal is mapped to a structured action with a logical action identifier, a configuration, and a target operation. The tested full configuration, B3, evaluates authorization before dispatch, keeps a durable reservation and in-flight state, and uses an adapter to communicate with a separately stored controlled target. The target maintains an effect oracle in its own SQLite store; this store is distinct from the runtime ledger, although both target and oracle are controlled by the study. A target record represents a committed target-side operation under the study's scope, not a verified downstream business loss.

\begin{figure}[t]
\centering
\resizebox{\columnwidth}{!}{%
\begin{tikzpicture}[
  node distance=10mm and 17mm,
  st/.style={draw, rounded corners=2pt, minimum width=26mm, minimum height=7mm, align=center, fill=gray!12, font=\small},
  obs/.style={draw, dashed, rounded corners=2pt, minimum width=26mm, minimum height=7mm, align=center, font=\small},
  lab/.style={font=\footnotesize, align=center},
  >=Stealth]
\node[st] (prop) {Proposed action};
\node[st, below=of prop] (res) {RESERVED};
\node[st, below=of res] (inf) {IN\_FLIGHT\\(target dispatch)};
\node[st, below=of inf] (comp) {COMPLETED};
\node[obs, right=of prop] (rej) {Rejected;\\no dispatch};
\node[obs, right=of inf] (unres) {Unresolved\\outcome};
\draw[->] (prop) -- node[lab, above] {invalid} (rej);
\draw[->] (prop) -- node[lab, left] {authorized:\\reserve} (res);
\draw[->] (res) -- node[lab, left] {mark\\in flight} (inf);
\draw[->] (inf) -- node[lab, left] {capture\\and commit} (comp);
\draw[->, dashed] (inf) -- node[lab, above] {no\\evidence} (unres);
\draw[->, dashed] (unres) |- node[lab, pos=0.3, right] {recovery: capture\\or target resolves\\outcome} (comp);
\end{tikzpicture}}
\caption{Simplified B3 action-record progression. Authorization is checked before a reservation is created. A repeated request for a completed action returns a deduplicated outcome without a new dispatch, and a repeated request for an unfinished action is reported as ambiguous. The unresolved outcome is an observation, not a persisted ledger state; if neither the capture nor target-side evidence establishes the outcome, the action remains unresolved.}
\label{fig:runtime}
\end{figure}
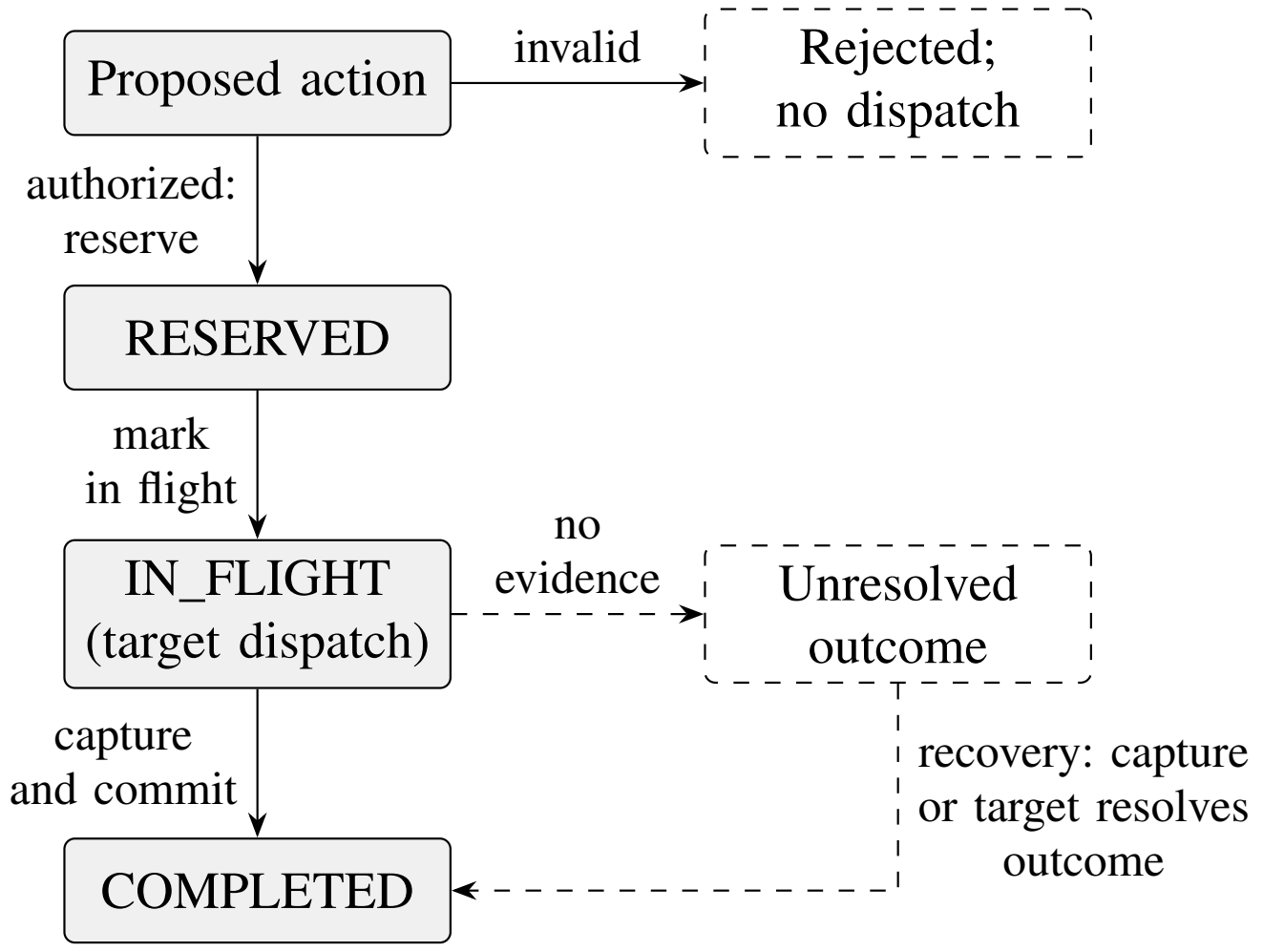

On an authorized action, B3 reserves an action record and marks it in flight before dispatch. A completed action can be suppressed on a repeated request; an existing unfinished action is returned as ambiguous rather than blindly treated as complete. Completion binds the ledger to a verified capture record. If capture persistence fails after the target effect, the completion step cannot establish that binding, so the action may remain in flight. Figure~\ref{fig:runtime} shows this distinction between a persisted state and an unresolved observation.

The runtime assigns a logical action identifier. For the tested formal driver, the target idempotency key is a SHA-256 digest over that identifier and the canonical payload digest; a separate invocation identifier incorporates the package, schedule, configuration, and action identity. The controlled target persists the key with the action scope and checks for conflicts in operation or payload. The key is not a secret and cannot compensate for a target that ignores or incorrectly implements the contract.

On recovery, an existing committed capture may support ledger completion. Otherwise, when reconciliation is enabled, the adapter consults target-side evidence and distinguishes an observed effect, an observed non-effect, and an unknown status. Unknown remains unresolved. A1 removes the reservation and completion bundle; A2 retains the ledger and target idempotency but disables the specified post-crash reconciliation. The A1 result is consistent with retained target-side controls contributing to duplicate suppression, but no separate intervention isolates the action identifier from target idempotency.

The capture journal hashes each body, binds the previous seal and action scope into metadata, and authenticates that metadata with HMAC-SHA-256. A committed manifest records the capture sequence; the ledger stores its capture binding. A3 removes sealing and full-history verification. The target, oracle, capture, and key are controlled test mechanisms. The evaluation does not establish protection against an adversary that controls all local copies or possesses the verification key.

\subsection{Configurations and ablations}
Table~\ref{tab:configs} summarizes the planned capability differences. B0 is a volatile fail-open baseline, B1 is volatile but rejects invalid authority, and B2 retains durable authorization with ordinary logging. B3 combines the capabilities above. The three simpler configurations and the ablations are internally constructed treatments of the same study, not independent commercial products.

\begin{table}[t]
\caption{Configuration roles in the controlled comparison}
\label{tab:configs}
\centering
\begin{tabular}{p{0.12\columnwidth}p{0.78\columnwidth}}
\toprule
Config. & Main capability or change \\
\midrule
B0 & Volatile, fail-open execution; no durable action reservation or replay suppression. \\
B1 & Fail-closed authorization, without durable action reservation or replay suppression. \\
B2 & Durable authorization and ordinary log; no durable action reservation or replay suppression. \\
B3 & Fail-closed authorization, durable reservation and completion, target idempotency, reconciliation, and sealed evidence. \\
A1 & B3 without reservation, in-flight record, completion transaction, and pre-dispatch replay suppression. \\
A2 & B3 without post-crash reconciliation. \\
A3 & B3 without evidence sealing and complete-history verification. \\
\bottomrule
\end{tabular}
\end{table}

A1 changes four related capabilities together; an A1--B3 difference cannot be attributed to any single one without a further isolating intervention. A2 retains the target idempotency and ledger capabilities while disabling the declared reconciliation capability. Consequently, a lack of repeated target effects in A2 cannot be used to infer that the ledger alone or the runtime alone provides a general guarantee.

\section{Fault Model}
The benchmark declares ten fault classes, summarized in Table~\ref{tab:faults}. The fault matrix defines admissible cutpoints and operations, rather than assigning every fault to every possible state. A fault's presence, an outcome, and an outcome's interpretability are recorded separately. In particular, an absent receipt is not converted into proof that a fault was exercised.

\begin{table}[t]
\caption{Declared fault classes and principal observation focus}
\label{tab:faults}
\centering
\setlength{\tabcolsep}{3pt}
\begin{tabular}{p{0.09\columnwidth}p{0.40\columnwidth}p{0.37\columnwidth}}
\toprule
ID & Injected condition & Observation focus \\
\midrule
F1 & Invalid or stale authority & Unauthorized effects \\
F2 & Crash after reservation & Lost effect, recovery \\
F3 & Crash in flight, before effect & Lost effect, recovery \\
F4 & Effect followed by response loss or delay & Duplicate, ambiguity, recovery \\
F5 & Effect before ledger completion, then crash & Orphaned effect \\
F6 & Capture write or synchronization error & Missing or corrupt evidence \\
F7 & Ledger completion failure & Orphaned effect, recovery \\
F8 & Pause, resume, or hard stop & False block, recovery \\
F9 & Replay or duplicate delivery & Duplicate effect \\
F10 & Evidence tampering & Detection, false alarm, localization \\
\bottomrule
\end{tabular}
\end{table}

The target implements three operation types: append an alert record (OP1), upsert a ticket (OP2), and transition a case state under a prior-state precondition (OP3). The controlled target is not a production email, financial, medical, or other external service. In F6 the harness injects an I/O exception at a capture boundary; no physical disk failure is claimed. F10 uses controlled evidence modifications and a test key; it is not a red-team evaluation against an adversary with unrestricted host or key access.

\section{Experimental Design and Analysis}
\subsection{Inputs and executions}
The study constructed 24 benchmark cases, with two cases for each combination of three operations and four declared strata. The recorded model interface aliases were \url{glm-5-3-flash} and \url{deepseek-v4-1-flash}; each produced two captured responses per case, giving 96 captured responses. These responses yielded 42 distinct normalized proposal digests. Model outputs were acquired before the fault runs and reused across configurations; no model or provider was invoked for each trial. For OP3, the initial target states were bound before execution. The names are recorded interface aliases, not proof of a specific internal model revision; the study does not compare the models.

On each platform, the original plan contained 4,000 clean-control trials, 2,250 primary-fault trials, 1,075 ablation trials, and 188 dedicated recovery trials, for 7,513 recorded trials. The dedicated recovery plan specified 200 planned restart operations per platform, with some trials accounting for more than one planned restart; planned restarts, observed restarts, and recovery successes are distinct quantities. The original analysis plan and source allocation were fixed before the physical fault trials, but after the captured responses already existed.

The Windows run used Python 3.12.13 on a local host with 24 logical processors, below-normal process priority, and output on an external USB volume. The pre-specified Linux platform was local WSL2; as a documented deviation, the Linux run used a four-vCPU, 16-GiB cloud virtual machine on an ext4 volume. The machines, storage, priorities, and background environments differ. Linux's monotonic and raw-clock diagnostics were recorded, but the short diagnostic is not an absolute long-duration clock calibration. Therefore, the platforms are reported separately and no direct hardware-performance ranking is inferred.

The original analysis plan is present in source commit \texttt{284c526d}, recorded at 2026-11-04 15:10:58 UTC; the Windows execution preflight recorded the plan hash at 2026-11-05 00:52 UTC, before that run. The supplementary analysis plan is present in commit \texttt{bd978393}, recorded at 2026-11-06 03:36:28 UTC; the supplementary runs started at 03:43 UTC on both platforms. These are repository and execution records rather than independently timestamped external registration. The original and supplementary plan hashes are retained in the study records.

\subsection{Outcome definitions and pairing}
The target oracle records target-side operations independently of the tested runtime ledger. A duplicate-effect outcome is defined as more than one recorded target effect associated with the logical action scope. An orphaned effect combines an observed target effect with the absence of a runtime ledger terminal under the study's metric definition. A recovery-success outcome requires the relevant declared recovery conditions; it is not defined for a configuration without a recovery path. Evidence missing, evidence corrupt, and unresolved runtime state are separate outcomes.

The principal contrasts pair configurations on the planned schedule, operation, source proposal, and platform. The original analysis plan declares 44 binary comparisons per platform across six families with sizes 3, 9, 14, 6, 5, and 7. Exact paired tests \cite{mcnemar1947note}, Newcombe paired intervals \cite{newcombe1998paired}, and within-family Holm adjustment \cite{holm1979simple} were specified; test probabilities are secondary to counts and effect estimates. A separately specified proposal-cluster bootstrap sensitivity resamples distinct normalized proposals and is reported without changing the primary test. Such clustering is important when observations share inputs, but it does not guarantee valid inference with few or dependent clusters \cite{cameron2015cluster}. Unknown and not-applicable observations remain separate.

Results are reported in three evidence layers. Layer P contains the original pre-specified analysis: the 44 paired comparisons per platform and the pre-specified descriptive metrics. Where every member of a family was observed, the within-family Holm adjustment was completed; where an applicable member remained unknown, the family is reported as incomplete, and its fully observed members are described by counts and intervals only. Layer S contains the F4 supplement, whose plan was fixed after the original outcomes were known but before its own trials; it is analyzed in its own families and does not alter Layer P. Layer H contains post-hoc case- and proposal-level direction summaries, sign calculations, multiplicity bounds, and the scoring replay; these describe robustness and are not confirmatory. Unless stated otherwise, counts are per platform, the two platforms are not pooled, and unknown, not-applicable, and failed observations are kept separate. These conventions apply throughout Section~\ref{sec:results} and are not repeated for each result. The full set of 44 pre-specified contrasts per platform, including missingness and interval fields, is retained in the study records.

Table~\ref{tab:metrics} defines the principal outcomes. The complete set of 33 metric definitions is retained in the study records.

\begin{table*}[t]
\caption{Outcome definitions and applicable faults}
\label{tab:metrics}
\centering
\setlength{\tabcolsep}{3pt}
\begin{tabular}{>{\raggedright\arraybackslash}p{0.19\textwidth}>{\raggedright\arraybackslash}p{0.52\textwidth}>{\raggedright\arraybackslash}p{0.17\textwidth}}
\toprule
Outcome & Counted when & Scope \\
\midrule
Unauthorized effect & The oracle observes an external effect despite invalid or missing authority & F1 \\
Duplicate effect & The oracle observes more than one external effect with the same logical action ID & F4, F5, F7, F9 \\
Lost effect & The runtime reports completion but the oracle observes no external effect & F2, F3 \\
Orphaned effect & The oracle observes an external effect but the ledger has no terminal record & F5, F7 \\
Ambiguous outcome & The runtime cannot determine whether an external effect occurred & F4, F5 \\
Unresolved state & The runtime remains in flight or unresolved after a recovery attempt & Storage and restart faults \\
Recovery success & A restart reaches the correct terminal state without operator intervention & F2--F4, F7, F8; recovery path required \\
Eventual success & The operation eventually succeeds after a transient fault or pause & F4, F8 \\
Correct effect once & The oracle observes exactly one valid external effect and the runtime completes & Applicable fault trials \\
Evidence missing & The terminal evidence record is absent from durable storage & F6; B3, A3 \\
Evidence corrupt & The evidence record fails checksum or cryptographic verification & F6; B3, A3 \\
Tamper detection & The verifier reports an injected evidence modification & F10; B3, A3 \\
Tamper false positive & The verifier flags an unmodified clean run & Clean B3 \\
Tamper localization & The verifier identifies the exact modified byte or record index & Detected F10 in B3 \\
Eligible success; false block & An authorized clean operation completes successfully; or is rejected by runtime policy & Clean B0--B3 \\
Recovery time & Wall-clock time from recovery invocation to terminal disposition & Successful recoveries \\
\bottomrule
\end{tabular}
\end{table*}

\subsection{F4 supplementary protocol}
The original F4 observations lacked direct receipts distinguishing response drop and delay. After inspecting those original outcomes, we fixed a separate F4 supplement before its new trials began. It used the same planned schedules and captured proposals, with added fault-harness receipts and corresponding verification; the tested runtime code was unchanged. This supplement examines the previously unknown fault class, but does not retroactively change the original observations. It is not a blinded holdout or an independent model sample. Its three F4-only multiple-comparison families used sizes 6, 3, and 4 and are not pooled with the original families. Three recovery comparisons against B0--B2 are structurally not applicable because those configurations do not implement a recovery path.

The supplement planned 500 trials per platform: 100 schedules for each of B0, B1, B2, B3, and A2. Linux recorded and recomputed all 500, with all 500 F4 stimuli verified. Windows recorded and recomputed 499. One A2 trial timed out during startup before the first dispatch and had no result; the remaining unstarted trials were subsequently run without repeating a started trial. The specific startup function and root cause were not retained. The missing observation remains unknown.

\subsection{Replay and source dependence}
The original planned 10\% scoring replay was performed by selecting 1,503 records after the original outcomes were available, using only plan fields and a fixed hash-derived rule: 750 on Windows and 753 on Linux. The original pre-outcome sampling seed was not located. The replay checked target records, effect counts, and applicable runtime verification using the existing scoring functions. It is a post-outcome-selected consistency check, not an independently implemented audit of all 33 metrics.

\section{Results}\label{sec:results}
\subsection{Original pre-specified comparisons}
Table~\ref{tab:fmea-results} connects observed failure modes with the mechanisms present in the compared configurations. Table~\ref{tab:main} reports selected original paired effects, trial-level intervals, and the proposal-cluster sensitivity. The study records retain all 44 pre-specified contrasts on each platform, including incomplete and non-estimable results; they are not deposited with this preprint.

\begin{table*}[t]
\caption{Failure modes, tested mechanisms, and selected observations. Counts are per platform; the F4 supplement is separate from the original analysis. Tags: [P] original pre-specified analysis; [S] pre-specified F4 supplement.}
\label{tab:fmea-results}
\centering
\begin{tabular}{p{0.17\textwidth}p{0.25\textwidth}p{0.46\textwidth}}
\toprule
Failure mode & Tested mechanism or contrast & Observation and evidence boundary \\
\midrule
Unauthorized effect & Fail-closed authorization & B0 100/100; B1--B3 0/100 in F1. [P] \\
Repeated effect & Identity and target idempotency retained in A1 & Original known F5/F9 pairs: B3 0/200, each B0--B2 117/200 [P]; F4 supplement: 0/100 vs.\ 66/100 [S]. \\
Orphaned effect & Ledger-related bundle & B3 0/170, A1 170/170; the ablation removes several capabilities together. [P] \\
Orphan or ambiguity & Post-crash reconciliation & F5 known pairs: B3 0/100, A2 100/100 for each outcome [P]. F4 supplement: A2 ambiguous 100/100 (Linux) and 99/99 (Windows), B3 0/100 and 0/99 [S]. \\
Recovery failure & Ledger-related bundle & B3 300/300 and A1 200/300 recovery successes; timing is reported separately. [P] \\
Tamper not detected & Sealing and history verification & B3 detected 89/89 known cases, A3 0/89; localization and unknowns are reported separately. [P] \\
Missing evidence & Capture completion boundary & B3 missing 75/100 in F6; A3 missing 50/75 in its applicable subset. [P] \\
Operational cost & Full configuration versus simpler baselines & Lower measured operation-window throughput and higher latency; see Table~\ref{tab:cost}. [P] \\
\bottomrule
\end{tabular}
\end{table*}

\begin{table*}[t]
\caption{Selected original paired effects. Differences are B3 minus comparator. Intervals are trial-level Newcombe and proposal-cluster sensitivities, respectively; a point interval reflects the observed bootstrap inputs, not perfect precision. All rows are Layer P. Values are for Windows; Linux values are identical except the recovery proposal-cluster interval, $[0.27,0.42]$.}
\label{tab:main}
\centering
\begin{tabular}{p{0.29\textwidth}ccccc}
\toprule
Outcome, comparator & B3 / comparator & Pairs & Difference & Paired 95\% CI & Proposal-cluster 95\% CI \\
\midrule
Unauthorized effect, B0 & 0/100 / 100/100 & 100 & $-1.00$ & $[-1.00,-0.95]$ & $[-1.00,-1.00]$ (13) \\
F5 orphan effect, B0 & 0/100 / 100/100 & 100 & $-1.00$ & $[-1.00,-0.95]$ & $[-1.00,-1.00]$ (21) \\
Orphan effect, A1 & 0/170 / 170/170 & 170 & $-1.00$ & $[-1.00,-0.97]$ & $[-1.00,-1.00]$ (21) \\
Recovery success, A1 & 300/300 / 200/300 & 300 & $+0.33$ & $[0.28,0.39]$ & $[0.27,0.43]$ (28) \\
\bottomrule
\end{tabular}
\end{table*}

The F1 result covers six cases and 13 proposal digests; B1 and B2 also recorded no unauthorized effects in these pairs. The 170-pair A1 orphan comparison covers 12 cases and 21 proposals. A1 removes four related capabilities, so this difference cannot be assigned to a single component. For the 300 A1 recovery pairs, B3 met the recovery criterion in all 300 and A1 in 200, across 18 cases and 28 proposals. The original F1 and A1 families yielded within-family adjusted trial-level results; the individual F5 orphan contrast was fully observed but its wider family remained incomplete. Among the six pre-specified A1 comparisons, A1 and B3 did not differ in lost effects (0/200 each; this metric covers the F2 and F3 pre-effect crash faults), ambiguous outcomes (0/100 each), unresolved runtime states (0/200 each), or repeated effects (0/300 each); the A1 differences were confined to orphaned effects and recovery.

In post-hoc grouping, the authorization comparison had a net direction favoring B3 for all 6 cases and 13 proposal digests; the A1 orphan comparison did so for all 12 cases and 21 proposal digests. No case or proposal digest had a net direction unfavorable to B3 in either of these two contrasts.

In F5, each B0--B2 baseline had an external effect without a ledger terminal in 100/100 known pairs, versus 0/100 for B3. This metric partly reflects the fact that B0 and B1 lack a durable action ledger, making the A1 and A2 contrasts more informative about mechanisms within B3. As a post-hoc sensitivity, for each individual F5 orphan contrast, the exact raw paired probability multiplied by the pre-specified family size nine is approximately $1.42\times10^{-29}$; under valid individual tests this is a conservative bound regardless of the other member probabilities; it is not an output of the incomplete family adjustment.

For original F5 and F9 repeated-effect comparisons, B3 had 0/200 and each B0--B2 baseline 117/200 among known pairs. Another 100 pairs per contrast were unknown because the original F4 stimulus could not be verified. For F5 ambiguity, B3 had 0/100 and each baseline 34/100 among known pairs. A2 and B3 both had 0/200 repeated effects in the known F5/F9 pairs.

The F5 A2--B3 comparison gives a further observation: in 100 known pairs per platform, A2 had an orphaned effect and an ambiguous outcome in 100/100, while B3 had 0/100 for each. These values are consistent with the tested reconciliation capability's role.

\subsection{Response loss and delay: separate F4 supplement}
Table~\ref{tab:f4ops} shows the operation-level repeated-effect counts in the F4 supplement. For each subtype and platform, B0, B1, and B2 each had 33 of 50 trials with more than one recorded target effect, while B3 had 0 of 50. Across both subtypes this is 66/100 for each simpler baseline and 0/100 for B3. The within-family adjusted trial-level probability for each duplicate contrast was $8.13\times10^{-20}$.

\begin{table}[t]
\caption{F4 supplement: trials with repeated target effects by operation (Layer S)}
\label{tab:f4ops}
\centering
\begin{tabular}{lrrr}
\toprule
Operation & Trials & Each of B0--B2 & B3 \\
\midrule
OP1 alert append & 16 & 16 & 0 \\
OP2 ticket upsert & 17 & 17 & 0 \\
OP3 guarded transition & 17 & 0 & 0 \\
\midrule
Total per subtype & 50 & 33 & 0 \\
\bottomrule
\end{tabular}

\vspace{2pt}
\footnotesize Counts are identical for both F4 subtypes and both platforms. In OP3, each baseline trial recorded one target effect followed by a precondition rejection.
\end{table}

The operation-level breakdown explains why the simpler baselines did not repeat effects in every F4 trial. Per subtype and baseline, all 16 OP1 trials and all 17 OP2 trials recorded repeated target operations, whereas each of the 17 OP3 trials recorded one target effect followed by a precondition rejection. Across the two subtypes, the 11 proposal digests without repeated baseline effects all mapped to OP3. The tested target checks the prior state before a state transition; after the first transition, an identical subsequent request can fail that precondition. This observation does not mean that B0--B2 implement general deduplication. For OP2, repeated upserts may leave the same final value while still producing multiple target-side operation records; our duplicate metric counts those records rather than estimating business harm.

The supplement's other pre-specified families show the same pattern. On each platform, every B0--B2 baseline left the outcome ambiguous in 100/100 F4 trials and reached eventual success in 0/100, whereas B3 had 0/100 ambiguous outcomes and 100/100 eventual successes (Holm-adjusted $p$ of $9.47\times10^{-30}$ and $4.73\times10^{-30}$, respectively). Because B0--B2 have no recovery path, part of this difference follows from their design, as in the F5 orphan comparison. The A2--B3 contrast is more informative, because A2 retains the ledger and target idempotency and disables only post-crash reconciliation. On Linux, A2 and B3 both had 0/100 repeated effects, but A2 was ambiguous in 100/100 trials and recovered or eventually succeeded in 0/100, whereas B3 was ambiguous in 0/100 and recovered and eventually succeeded in 100/100 (Holm-adjusted $p=6.31\times10^{-30}$ for each of these three contrasts). On Windows, the 99 observed A2 pairs showed the same pattern (A2 ambiguous in 99/99; B3 recovered and eventually succeeded in 99/99); one A2 trial is missing, so the Windows A2 family has no completed adjustment. Duplicate suppression and resolution of an uncertain outcome are therefore separate properties: the first was present without reconciliation, the second was not.

An independent local SQL script checked logical target-effect counts across the original and supplementary F4 records: the counts matched in all 500 comparable Linux pairs and 499 comparable Windows pairs. The missing Windows trial was not compared. This supports agreement of the effect counts, not equality of all runtime behavior. The added fault receipts are sent and recorded while a fault is being handled, so they may affect timing or scheduling even though the final receipt drain occurs after the runtime stops.

\subsection{Evidence integrity and adverse outcomes}
For the declared F10 tamper comparison, each platform had 89 known B3/A3 pairs: B3 detected 89/89 and A3 0/89. Eleven further pairs were unknown or not triggered under the relevant controlled condition and were not added to the denominator. For the separate localization metric, 67 of 89 selected B3 observations were confirmed correctly localized and 22 were unknown; these 22 are not localization failures. In clean B3 trials, the observed tamper false-positive count was 0/1,000 per platform.

For the selected F7 effect comparisons, A3 and B3 both recorded 0/100 repeated-effect outcomes and 0/70 orphaned-effect outcomes. No effect-count difference was observed in those comparisons.

F6 exposes a material limitation. Per platform, 75/100 B3 trials had missing evidence and an unresolved runtime state, while 25/100 did not exhibit those two observations; the recorded corrupt-evidence count was 0/100. In the 75 F6 pairs applicable to A3, both A3 and B3 had 50 missing-evidence observations. A3 lacks the seal-specific boundary, so its 75-trial denominator is not interchangeable with B3's 100.

The common A3--B3 F6 pairs include 25 body-write-boundary trials without missing evidence under the study's metric, and 25 file-synchronization plus 25 directory-synchronization boundary trials with missing evidence in both configurations. B3 has a further 25 seal-write-boundary trials; combining the pre-specified common-pair counts with the B3 total implies 25/25 missing in that additional group. The implementation calls capture persistence after target dispatch and requires a suitable capture before committing ledger completion. An injected capture exception can therefore leave the action unfinished. This is consistent with declining to finalize without the required evidence, at an availability cost.

\subsection{Availability and operation-window overhead}
For each platform, the clean-control set included 1,000 operations per baseline configuration. All four configurations had 1,000/1,000 eligible successes and 0/1,000 false blocks in these controlled clean trials. B3's 1,000 trials reused only six cases and eight proposal digests, so this is not a 1,000-case estimate of field availability. Across the nine pre-specified B3--B1 and B3--B2 binary contrasts available on each platform, the B1 and B2 pair-count fields were identical.

Table~\ref{tab:cost} reports the study's executor operation-window throughput, which includes observed failed or blocked operations in its defined window and is not end-to-end product throughput. On Windows, the difference between B3 and B0 in their empirical latency quantiles was approximately $+91$ ms at p50, $+1.32$ s at p95, and $+1.49$ s at p99. These are differences of arm quantiles, not quantiles of per-pair latency differences. The median paired SQLite transaction-count difference was $+4$ relative to B0 and $+3$ relative to B2; the measured median storage difference versus B0 was 10,204 bytes per action.

\begin{table}[t]
\caption{Descriptive operation-window throughput in clean trials, operations per second. Hosts, storage, and process priority differ between platforms.}
\label{tab:cost}
\centering
\begin{tabular}{lrrrr}
\toprule
Platform & B0 & B1 & B2 & B3 \\
\midrule
Windows & 20.57 & 21.50 & 11.65 & 3.00 \\
Linux & 318.81 & 319.16 & 163.21 & 41.94 \\
\bottomrule
\end{tabular}
\end{table}

In the dedicated recovery family, 138/138 applicable recovery outcomes succeeded on each platform. Their median and nearest-rank p95 recovery times were 1.09 s and 2.13 s on Windows, and 0.28 s and 0.56 s on Linux. The 138 applicable outcomes are not the 188 planned recovery trials or 200 planned restart operations.

For the non-F1 correct-effect-once descriptive comparison, B3 met the criterion in 175/200 known pairs and B0 in 49/200; a further 100 pairs were eligibility-unknown. In 25 known pairs B0 met the criterion while B3 did not. Saved effect counts show one B3 target effect in each of those 25, but B3 did not complete its runtime outcome. The metric therefore captures completion as well as the target-effect count, and cannot be interpreted as a pure duplicate-rate measure.

\section{Design Implications and Answers to the Research Questions}
For designers of agent runtimes, the results translate into four implications within the tested boundary.

\emph{Fail closed before dispatch.} Rejecting invalid authority blocked every injected unauthorized action, and the volatile fail-closed configuration did so with operation-window throughput matching the fail-open baseline (Table~\ref{tab:cost}). In this test, B1 achieved the same authorization outcome without the full runtime: B1, B2, and B3 recorded identical counts on this outcome.

\emph{Give every action a stable identity and make the target honor it.} In the selected known original comparisons, repeated effects were absent in configurations that retained stable action identities and target-side idempotency, including A1 and A2. The experiment does not separate the identifier from the target's idempotency check, and the protection depends on a target that implements that contract.

\emph{Do not mistake idempotency for recovery.} Duplicate suppression did not make outcomes knowable. Removing the ledger bundle orphaned effects and reduced recovery; removing only reconciliation left uncertain outcomes ambiguous in both F5 and the F4 supplement. The tested runtime records intent before dispatch and consults the target after a failure; it does not infer completion from an unanswered call. Because A1 removes several ledger capabilities together, the experiment cannot tell which of them is individually necessary.

\emph{Budget for durability, and decide how to fail.} In the measured operation windows, the volatile fail-closed configuration had no lower throughput than the fail-open baseline. The durable-authorization and logging configuration had roughly half the measured throughput of B1, while their pair-count fields matched in nine comparisons against B3, and B3 ran at about one-seventh of fail-open throughput, with a median of four additional SQLite transactions and about 10~kB of additional storage per action. Timing was not isolated by component, so the share due to each mechanism, USB storage, scheduling, or instrumentation is unknown. Requiring evidence before completion also has an availability price: when capture persistence failed, B3 left 75/100 actions unresolved rather than finalize them without evidence. Whether to stop or to proceed in that situation is a policy decision that a runtime should make explicit.

\begin{table*}[t]
\caption{Bounded answers to the pre-specified research questions.}
\label{tab:rq}
\centering
\begin{tabular}{p{0.10\textwidth}p{0.77\textwidth}}
\toprule
Question & Answer supported by this study \\
\midrule
RQ1 & Fail-closed configurations blocked the measured unauthorized effects (0/100 versus 100/100); the result is not unique to B3, and the volatile fail-closed configuration did not have lower measured operation-window throughput than fail-open. \\
RQ2 & The full configuration had fewer repeated and orphaned effects in selected known pairs; disabling reconciliation alone left F5 and F4 outcomes ambiguous. Unknown original F4 stimuli and same-team target assumptions limit the claim. \\
RQ3 & The tested sealing/history configuration detected 89/89 known F10 changes versus 0/89 for A3; localization and other observations retain unknowns. \\
RQ4 & B3 recorded 300/300 recovery successes versus 200/300 for A1, and 138/138 in the dedicated recovery family (median 1.09 s on Windows, 0.28 s on Linux); capture failures and measured latency and throughput show availability and cost limitations. \\
RQ5 & All 44 pre-specified contrasts had identical pair-count fields on the two platforms. Both executions used the same captured proposals and fault classes, so this shows consistent behavior of one implementation on two hosts, not independent replication; timing differed, and a cloud host replaced the planned WSL2 environment. \\
\bottomrule
\end{tabular}
\end{table*}

As Table~\ref{tab:rq} indicates, RQ5 has a narrower answer than equality of all measurements. The 44 pre-specified contrast records have the same planned, known, unknown, and binary paired-count fields on Windows and Linux. The supplementary Windows experiment retained one missing A2 observation, and performance and recovery times differ. The shared inputs and implementation make these two executions a cross-platform consistency observation, not independent replications with newly sampled cases.

\section{Threats to Validity}\label{sec:threats}
\subsection{Dependence and inferential unit}
The 7,513 trials per platform reuse 24 cases and 96 responses, and the F4 supplement reuses 12 cases and 21 proposal digests. A paired-trial exact test does not make those repetitions independent new cases. In a post-hoc descriptive aggregation of F4 B3 versus B0, 8 of 12 cases had a net direction favoring B3 and 4 tied; 10 of 21 proposal digests favored B3 and 11 tied. No case or proposal digest had a net direction favoring B0 in this comparison.

As an illustration only, if the eight non-tied case-level signs were independent and exchangeable under a null hypothesis, a two-sided sign calculation gives $p=0.0078125$; a 13-comparison Bonferroni bound is approximately 0.102. The aggregation and test were selected after observing outcomes, and the independence and exchangeability assumptions were not established. This sensitivity neither proves that the measured effect is absent nor replaces the pre-specified trial-level analysis. It shows why very small trial-level $p$-values should not be read as evidence from hundreds of independent cases.

\subsection{Missing observations and measurement}
In the original analysis, each platform had 522 fault-stimulus states classified unknown: 500 original F4 observations and 22 F10 observations. The new F4 supplement is reported separately and does not rewrite those earlier states. Additional F9 and F8 conditions and per-operation synchronization counts retain the unknown status specified by the available evidence. An absent recovery path is not a failed recovery and does not remove a baseline from unrelated comparisons.

The Windows F4 supplement has one missing A2 result, so the Windows A2 supplementary family has no completed adjustment; the 99 observed A2 pairs showed the same pattern as the Linux pairs. The added receipts may perturb timing. The target oracle is separated from the runtime ledger but is implemented by the same research team and exercised in a controlled local environment.

\subsection{External validity and reproducibility}
The configurations are same-team implementations and ablations, not external competing frameworks. The experiment does not establish multi-host behavior, arbitrary concurrency, real provider outage rates, production authorization security, or live deployment reliability. The controlled target's preconditions and idempotency support are part of the causal setting. The two model aliases are descriptive source strata, not randomized model treatments.

The post-outcome 10\% replay found no disagreement in the compared target-record, effect-count, and applicable runtime-verification fields for 750 Windows and 753 Linux records. Some configurations did not have an applicable runtime verifier. Because the replay reused the scoring implementation and did not cover every metric, it cannot prove the scoring program correct or all stored data immutable. A separate SQL recount agreed with the saved target-effect facts for 2,750 F1, F5, and F7 records across the platforms; ledger completion was independently checked only where the corresponding ledger existed. Authorization classification remained shared with the original analysis.

\section{Conclusion}
An agent's tool call can change an external system without the agent knowing. Under ten classes of injected single-host faults, we compared how safeguard configurations and ablation bundles relate to target-side failures. Fail-closed configurations blocked the tested unauthorized effects; the volatile fail-closed configuration did not have lower measured operation-window throughput than fail-open. In selected known pairs, configurations retaining stable action identities and target-side idempotency recorded no duplicates, but this did not make outcomes recoverable: removing the ledger-related bundle coincided with orphaned effects and lower recovery, while removing reconciliation left outcomes ambiguous. A supplement designed after the original outcomes and reusing the same inputs showed these patterns under verified response loss or delay.

These benefits have a price. The full runtime's operation-window throughput was roughly one-seventh of the fail-open baseline's, and requiring evidence before completion left actions unresolved when evidence could not be written. The study is bounded by one runtime, a controlled target, reused inputs, and single-host faults, and its two platforms show consistent behavior of one implementation rather than independent replication. Future work should test independently sampled cases, production targets, and multi-host deployments, and should examine policies that decide how an agent proceeds when evidence cannot be recorded.

\section*{Acknowledgment}
The author used OpenAI Codex and Anthropic Claude (Opus) to assist with drafting and revising the manuscript, including the abstract, introduction, and design implications, and with preparing some analysis code.

\section*{Patent Disclosure}
The author is the inventor and rights holder of Chinese patent CN 121116528 B, which describes a related agent-collaboration architecture.

\section*{Data and Code Availability}
Only the manuscript is deposited with this preprint. The aggregate analysis tables, experimental code, full original records, and execution environment are not publicly deposited with it; access is subject to author authorization and privacy review.

\bibliographystyle{IEEEtran}
\bibliography{references}
\IfFileExists{biographies.tex}{\input{biographies}}{}
\end{document}